\documentstyle[12pt]{article}
\begin{document}
\input FEYNMAN
\input FERMIONSETUP
\input GLUONSETUP
\newcommand{\beq}{\begin{equation}}
\newcommand{\eeq}{\end{equation}}
\newcommand{\bea}{\begin{eqnarray}}
\newcommand{\eea}{\end{eqnarray}}
\newcommand{\dirac}{/\!\!\!\partial}
\newcommand{\Dirac}{/\!\!\!\!D}
\def\lequiv{\raise 0.4ex \hbox{$<$} \kern -0.8 em \lower 0.62 ex \hbox{$\sim$}
}
\def\gequiv{\raise 0.4ex \hbox{$>$} \kern -0.7 em \lower 0.62 ex \hbox{$\sim$}
}
\begin{titlepage}
\begin{center}
\hfill gr-qc/9301012  \\
\hfill NYU-TH.93/01/01
\vskip .4in
{\large\bf Massive Spin-5/2 Fields Coupled to Gravity:
Tree-Level Unitarity vs. the Equivalence Principle}
\end{center}
\vskip .4in
\begin{center}
{Massimo Porrati\footnotemark}
\vskip .4in
{\it Department of Physics, New York University\\
4 Washington Pl., New York NY 10003, USA.}
\footnotetext{On leave of absence from I.N.F.N., sez. di Pisa, Pisa, Italy.}
\end{center}
\vskip .4in
\begin{center} {\bf ABSTRACT} \end{center}
\begin{quotation}
\noindent
I show that the gravitational scattering amplitudes of a spin-5/2
field with mass $m\ll M_{Pl}$ violate tree-level unitarity at energies
$\sqrt{s}\approx\sqrt{mM_{Pl}}$ if the coupling to gravity is minimal.
Unitarity up to energies $\sqrt{s}\approx M_{Pl}$ is restored by adding a
suitable non-minimal term, which gives rise to interactions violating the
(strong) equivalence principle. These interactions are only relevant
at distances $d\lequiv 1/m$.
\end{quotation}
\vfill
\end{titlepage}
\noindent
Torsion-free minimal coupling to
gravity is defined by the following substitutions, performed on a
free lagrangian for an arbitrary field $\phi$:
\beq
\partial_{\mu}\rightarrow D_{\mu}, \;\;\; \eta^{\mu\nu}\rightarrow
g^{\mu\nu}=e^{\mu}_a e^{\nu\, a},
\label{1}
\eeq
and by the equation $D_{\mu}e^a_{\nu}=0$.
The flat metric $\eta^{\mu\nu}$ is replaced by the curved one, and ordinary
derivatives are replaced by covariant derivatives, containing the usual
symmetric Christoffel symbol, and the spin-connection.
The tetrads $e^a_{\mu}$, which
transform curved indices $\mu$ into flat tangent-space indices $a$, are
needed in order to describe the coupling of fermions to gravity.

The equivalence principle, in its weakest form, states that gravity
couples to the stress-energy tensor of matter $T_{\mu\nu}$.
To linear order in the fluctuations of the
gravitational field about the flat-space background, and in the gauge
$e_{a\mu}-e_{\mu a}=0$,
\beq
S[\phi, e^a_\mu]= \left.S[\phi]\right|_{free} + T^{\mu\nu}[\phi]h_{\mu\nu},
\;\;\; h_{\mu\nu}=e_{\mu\nu}-\eta_{\mu\nu}.
\label{2}
\eeq
Here $S[\phi, e^a_\mu]$ is the action of the field $\phi$ coupled to gravity.
Notice that at linear order we may identify curved- and flat-space indices.
In this formulation the equivalence principle is simply a definition of the
stress-energy tensor, through eq.~(\ref{2}).

The minimal coupling prescription is singled out by imposing that $T_{\mu\nu}$
be the (symmetrized) Noether stress-energy tensor.
Needless to say, the minimal-coupling prescription is neither unambiguous nor
unique: covariant derivatives do not commute, therefore, integrating by parts
in a free bosonic lagrangian, and using prescription~(\ref{1}) one finds a
different result than by performing substitution~(\ref{1}) directly. Moreover,
general relativity allows for matter-gravity couplings proportional
to the curvature tensors, and derivatives thereof. In addition to those
problems, massless fields coupled to gravity may turn out to give inconsistent
theories~\cite{AD1}. In the following, I shall deal with massive fields only,
where this last problem does not arise.

The former two problems of minimal coupling are of little concern in the case
of massive particles of spin s~$\leq 2$ interacting with gravity.
The reason behind this fact is that the minimal coupling of particles with
s~$\leq 2$, besides being ``simple,'' enjoys another remarkable property:
scattering amplitudes, involving gravitational interactions only, are small for
any center-of-mass energy $\sqrt{s}$ lower than the Planck scale $M_{Pl}$.

In order to convince ourselves that this property is not obvious, let us
consider, for instance, the conversion of two massive, spin-s particles into
two gravitons. The graphs contributing to this
process are depicted in fig.~1. The corresponding scattering amplitude
involves the propagator $\Pi$ of the massive field $\phi$.\\
\begin{picture}(100000,16500)
\drawline\fermion[\SE\REG](0,15500)[5000]
\drawline\gluon[\NE\REG](\particlebackx,\particlebacky)[3]
\drawline\fermion[\S\REG](\gluonfrontx,\gluonfronty)[7500]
\drawline\gluon[\SE\REG](\particlebackx,\particlebacky)[3]
\drawline\fermion[\SW\REG](\gluonfrontx,\gluonfronty)[5000]
\drawline\fermion[\SE\REG](19000,15500)[5000]
\drawline\gluon[\SE\REG](\particlebackx,\particlebacky)[6]
\drawline\fermion[\S\REG](\gluonfrontx,\gluonfronty)[7500]
\drawline\gluon[\NE\REG](\particlebackx,\particlebacky)[6]
\drawline\fermion[\SW\REG](\gluonfrontx,\gluonfronty)[5000]
\drawline\fermion[\NW\REG](41500,8550)[6000]
\drawline\gluon[\NE\REG](\pfrontx,\pfronty)[4]
\drawline\gluon[\SE\REG](\particlefrontx,\particlefronty)[4]
\drawline\fermion[\SW\REG](\particlefrontx,\particlefronty)[6000]
\end{picture}
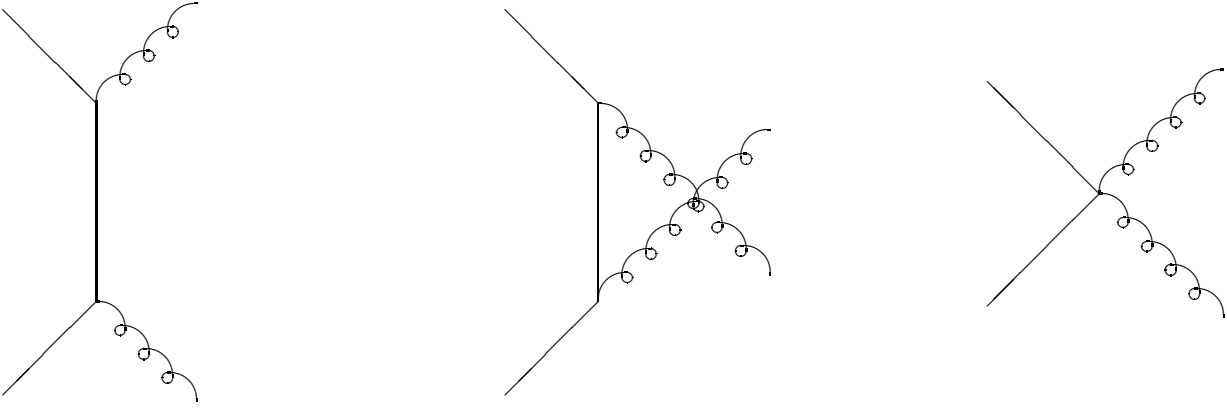
Figure 1: solid lines denote massive, spin-s particles, curly lines denote
gravitons.
\vskip .1in
It is well
known that for s~$\geq 1$ this propagator contains terms proportional to
$1/m^2$, due to the existence of (restricted) gauge invariances in the
$m\rightarrow 0$ limit. These
mass singularities could, in principle, give rise to a scattering amplitude
containing terms ${\cal O}(s^2/m^2M_{Pl}^2)$. Such a scattering amplitude
would become large, and eventually exceed the unitarity bounds, at
$\sqrt{s} \approx \sqrt{m M_{Pl}}$.
This is an energy scale much below the Planck one, when $m\ll M_{Pl}$.

The reason why ${\cal O}(s^2/m^2M^2_{Pl})$ terms are absent for s=1, 3/2, 2
is the following.
The diagrams in fig.~1 giving rise to the dangerous ${\cal O}(s^2/m^2M_{Pl}^2)$
terms have the form $J\Pi J$\footnotemark.
\footnotetext{The `seagull'' diagram in fig.~1 does not contribute to the
leading zero-mass singularity.}
The tensor current $J$ is
obtained by varying the action $S[\phi, e^a_{\mu}]$ with respect to the field
$\phi$, and keeping only terms linear in the fluctuation of the metric about
the flat-space background
\beq
{\delta S[\phi, e^a_\mu]\over \delta \phi} = J+ {\cal O}(h^2).
\label{3}
\eeq
As noticed above, terms proportional to $1/m^2$ in the propagator $\Pi$ are
related to gauge invariances in the massless limit. More precisely
$\Pi\stackrel{\,\,\,o}{\phi}=m^{-2}\stackrel{\,\,\,o}{\phi}$ iff
$\stackrel{\,\,\,o}{\phi}$ is a pure gauge. The
standard form of $\stackrel{\,\,\,o}{\phi}$ for s=1, 3/2, 2 reads\footnotemark
\footnotetext{More complicated forms of $\stackrel{\,\,\,o}{\phi}$ can be
reduced
to eq.~(\ref{4}) by field redefinitions.}
\beq
{\rm s}=1:\; \stackrel{\,\,\,o}{\phi}_\mu=\partial_\mu \epsilon,\;\;
{\rm s}=3/2:\; \stackrel{\,\,\,o}{\phi}_\mu=\partial_\mu \epsilon,\;\;
{\rm s}=2:\; \stackrel{\,\,\,o}{\phi}_{\mu\nu}=\partial_\mu \epsilon_\nu
+\partial_\nu\epsilon_\mu.
\label{4}
\eeq
The gauge parameter $\epsilon$ is a real scalar for s=1, a
Majorana spinor for s=3/2, and a real vector for s=2.
If the projection $J\cdot\stackrel{\,\,\,o}{\phi}$
of the current $J$ on the vectors $\stackrel{\,\,\,o}{\phi}$ has the
form $mX$, with $X$ any operator possessing a smooth $m\rightarrow 0$ limit,
then, by dimensional reasons, no ${\cal O}(s^2/m^2M^2_{Pl})$ terms
will arise in
the scattering amplitude of fig.~1. The key observation now is that, up to
${\cal O}(h^2)$ terms, $J\cdot\stackrel{\,\,\,o}{\phi}$ equals
$\stackrel{\,\,\,o}
{\phi}\cdot\delta S[\phi,e^a_{\mu}]/\delta\phi$, due to eq.~(\ref{3}):
we find the projection $J\cdot\stackrel{\,\,\,o}
{\phi}$  by varying the action
$S[\phi,e^a_{\mu}]$ under a gauge transformation\footnotemark,
\footnotetext{To be exact, under a gauge transformation of the massless,
free lagrangian
$S_0[\phi]=\lim_{m\rightarrow 0}S[\phi]$.}
and linearizing in the
gravitational field $h_{\mu\nu}$.

For generic spin this variation contains terms of the form $mX$, hereafter
called ``soft,'' as well as hard
terms. The latter ones do not vanish in the $m\rightarrow 0$ limit.
For s=1, 3/2, 2 though, the hard terms are proportional either to the
(linearized)
scalar-curvature tensor $R$ or to the Ricci tensor $R_{\mu\nu}$~\cite{DZ,AD2}.
These hard contributions vanish when we impose the free-graviton equations of
motion (that is the linearized Einstein equations in the vacuum).
We are allowed to use Einstein's equations because the graviton lines in fig.~1
are external.
Moreover, ambiguities in the ordering of covariant derivative,
in the action of spin-1 and
spin-2 fields, yield only harmless terms, vanishing on shell.

Up to spin 2 minimal coupling seems therefore the only natural choice for
describing
gravitational interactions, but the situation changes drastically for
s=5/2.

Lagrangians for massive particles of spin 5/2 have been given by several
authors~\cite{AD1,B,S}, we adopt here the lagrangian given in
refs.~\cite{B,PvN}.
The set of fields used there is the minimal one needed to describe a spin-5/2
particle~\cite{SH,B}, namely, a Majorana tensor-spinor $\psi_{\mu\nu}$,
symmetric in the $\mu$, $\nu$ indices, and an auxiliary Majorana spinor
$\chi$. The action reads
\bea
S&=& \int d^4x e [ -{1\over 2} \bar{\psi}_{ab}\Dirac\psi_{ab} -\bar{\psi}_{ab}
\gamma_b \Dirac \gamma_c\psi_{ca} + 2\bar{\psi}_{ab}\gamma_bD_c\psi_{ca}
 +{1\over 4} \bar{\psi}_{aa}\Dirac \psi_{bb}
-\bar{\psi}_{aa}D_b\gamma_c\psi_{bc} ]
\nonumber \\
& & + {m\over 2} ( \bar{\psi}_{ab}\psi_{ab}
 -{3\over 4} \bar{\psi}_{ab}\gamma_b \gamma_c\psi_{ca} -{7\over 4}
\bar{\psi}_{aa}\psi_{bb} -{16\over 3} \bar{\chi}\psi_{aa} -{32\over
9}\bar{\chi}\chi).
\label{5}
\eea
Our conventions on the metric and gamma matrices follow ref.~\cite{PvN}.
The covariant derivative of the field $\psi_{\mu\nu}$ is
\beq
D_\mu\psi_{\nu\rho}=\partial_\mu\psi_{\nu\rho} + {1\over 2}
\sigma_{ab}\omega_{\mu}^{ab}(e)\psi_{\nu\rho} +
\Gamma^\lambda_{\mu\nu}\psi_{\lambda\rho}
+\Gamma^\lambda_{\mu\rho}\psi_{\nu\lambda}.
\label{6}
\eeq
The Christoffel's symbols are the standard ones (torsion-free). The commutator
of two covariant derivatives is
\beq
[D_\mu, D_\nu]\psi_{\rho\sigma}
 ={1\over 2} R_{\mu\nu ab}\sigma^{ab}\psi_{\rho\sigma} +
               R_{\mu\nu\rho\lambda}\psi^{\lambda}_\sigma +
R_{\mu\nu\sigma\lambda}\psi^{\lambda}_\rho.
\label{7}
\eeq
The free lagrangian possesses a restricted gauge invariance
at $m=0$~\cite{B,S}.
The gauge parameter is a gamma-traceless Majorana vector spinor, and the
gauge transformation reads
\beq
\delta\psi_{\mu\nu}=\partial_\mu\epsilon_\nu + \partial_\nu\epsilon_\mu,\;\;\;
\gamma^\mu\epsilon_\mu=0.
\label{8}
\eeq
The free equations of motion of $\psi_{\mu\nu}$ and $\chi$ are
\beq
\dirac\psi_{\mu\nu}=m\psi_{\mu\nu},\;\;
\gamma^\mu\psi_\mu=\partial^\mu\psi_{\mu\nu}=0,\;\; \chi=0.
\label{9}
\eeq

In order to see whether ${\cal O}(s^2/m^2M_{Pl}^2)$ terms exist in the
scattering diagrams of fig.~1 we must perform a variation of action~(\ref{5})
under the transformation~(\ref{8}), linearize in the gravitational field, and
put $\psi_{\mu\nu}$, $\chi$ and $h_{\mu\nu}$ on shell. A short calculation
gives
\beq
\delta
S=-4\bar{\epsilon}_\nu\gamma_\rho\psi_{\lambda\sigma}R^{\nu\lambda\rho\sigma}
+ {\rm soft\; terms} + {\cal O}(h^2).
\label{10}
\eeq
The hard term in this equation is proportional to the Riemann
tensor, thus, it does not vanish on shell, and the scattering amplitude of
fig.~1 does contain ${\cal O}(s^2/m^2M_{Pl}^2)$ terms.

This result means that a minimally coupled light ($m\ll M_{Pl}$) spin-5/2 field
interacts strongly with gravity even at relatively low energies
($\sqrt{s}\approx \sqrt{mM_{Pl}}\ll M_{Pl}$)\footnotemark.
\footnotetext{A massless spin-5/2 field coupled to gravity is downright
inconsistent~\cite{AD1}.}
This scenario seems bizarre: it seems natural to assume that gravitational
interactions be weak up to energies
$\sqrt{s}\approx M_{Pl}$, irrespective of any particle's mass. If we do impose
this requirement, or, in other words, if we impose that gravitational
tree-level amplitudes respect unitarity up to the Planck scale, we must find a
way of cancelling the hard term in eq.~(\ref{10}), even at the price of giving
up the minimal-coupling prescription. Minimal coupling has nothing really
fundamental about it, whereas tree-level unitarity is a sensible requirement
which, in the case of electromagnetic interactions, has been already
proven fruitful~\cite{W,LS,FPT}.

Indeed, we can cancel the hard term of eq.~(\ref{10}) by adding to the
spin-5/2 action a non-minimal
coupling proportional to the Riemann tensor.
Notice that a similar situation happens when higher-spin massive particles are
coupled to electromagnetism~\cite{FPT}. In that case,
appropriate non-minimal terms
cancel tree-level unitarity violating terms in the scattering
amplitudes~\cite{FPT}. In ref.~\cite{FPT} it was also shown that the
cancellation between minimal and non-minimal terms does occur
for charged open-string states in a constant
electromagnetic background.

To prove that the cancellation takes place also for a spin-5/2 field coupled
to gravity,
let me add a non-minimal term to action~(\ref{5})
\beq
\int d^4x e \alpha\bar{\psi}_{\mu\nu}(R^{\mu\rho\nu\sigma}
+{1\over 2}
\gamma^5\epsilon^{\nu\sigma\alpha\beta}R^{\mu\rho}_{\alpha\beta})
\psi_{\rho\sigma}.
\label{11}
\eeq
The variation of this term under transformation~(\ref{8}) yields
\bea
-2\alpha\bar{\epsilon}_\mu
R^{\mu\rho\nu\sigma}(\partial_\nu\psi_{\sigma\rho}
 -\partial_\sigma\psi_{\nu\rho}
+\gamma^5\epsilon_{\nu\sigma}^{\;\;\;\;\alpha\beta}
\partial_\alpha\psi_{\beta\rho})
&=& 2\alpha m\bar{\epsilon}_\mu
R^{\mu\rho\nu\sigma}\gamma^\lambda\sigma_{\nu\sigma}\psi_{\lambda\rho}
+ {\cal O}(h^2). \nonumber \\
& & \label{12}
\eea
To get eq.~(\ref{12}) one must recall that, on shell
\beq
D_\mu R^{\mu\nu\rho\sigma}=0,\;\;\; \dirac\psi_{\mu\nu}=m\psi_{\mu\nu},
\label{13}
\eeq
use the Bianchi identitites
\beq
R^\mu_{[\nu\rho\sigma]}=0,\;\;\; \partial_{[\mu}R_{\nu]\rho\sigma\tau}=0,
\label{14}
\eeq
and the identity $R^{\alpha\beta\gamma[\mu}\epsilon^{\nu\rho\alpha\beta]}=0$.
Recalling that on shell $\gamma^\mu\psi_{\mu\nu}=0$, eq.~(\ref{12}) is
transformed into
\beq
-4\alpha m \bar{\epsilon}_\nu
R^{\nu\lambda\rho\sigma}\gamma_\rho\psi_{\lambda\sigma}.
\label{15}
\eeq
By choosing $\alpha=-1/m$ the term~(\ref{15}) cancels the hard term in
eq.~(\ref{10}), and, by consequence,
the terms proportional to $s^2/m^2M_{Pl}^2$ in the
scattering amplitude of fig.~1.

Notice that our cancellation becomes exact on {\em any} background
satisfying Einstein's vacuum equations, if we covariantize the derivatives
in eq.~(\ref{8}). The coupling~(\ref{11}), needless to say, is defined only up
to terms proportional to $R$ and $R_{\mu\nu}$.

The non-minimal term in eq.~(\ref{11}) violates the strong
equivalence principle, and it introduce a coupling proportional
to $1/m$. The new stress-energy tensor associated with the lagrangian
containing~(\ref{11}) is no longer the Noether one; for instance, it
contains terms with two derivatives of the field. There exist a field
redefinition transforming this new stress-energy tensor into the standard one,
but it is probably non local, being defined as a power series in $1/m$.
This new coupling is obviously compatible with all principles of
general relativity. Moreover, it is negligible at large distance, i.e. in
soft-graviton scattering with transferred momenta
$q\lequiv m$, since it gives rise to additional interactions
of strength proportional to $q^2/m$. This is to be compared with the
minimal interaction, whose strength is proportional to the energy $E$
of the spin-5/2 particle.

Tree-level unitarity up to $\sqrt{s}=M_{Pl}$ seems a physically
meaningful and natural requirement, unlike the minimal-coupling prescription.
This requirement entails that any theory containing light spin-5/2 particles
should give rise to an effective lagrangian containing the term~(\ref{11})
with $\alpha=-1/m$.
String theory is an example of such a theory,
when the string tension $\alpha'\ll M_{Pl}^{-2}$:
it would be interesting to check whether its low-energy effective lagrangian
actually contains term~(\ref{11}).

\end{document}